\documentclass[aps,prl,twocolumn,superscriptaddress,a4paper,english,longbibliography]{revtex4-1}

\usepackage{times}
\usepackage{color}
\usepackage[colorlinks=true,urlcolor=blue,citecolor=blue]{hyperref}
\usepackage{url}
\usepackage{breakurl}
\usepackage{soul}
\usepackage{graphicx}
\usepackage{amsmath}
\usepackage{subfigure}
\usepackage{xcolor}
\usepackage{amssymb}
\usepackage{latexsym}
\usepackage{hyperref}% add hypertext capabilities
\DeclareGraphicsExtensions{.png,.eps}
\usepackage{float}
\usepackage{ulem}
\usepackage{appendix}
\usepackage{etoolbox}

\usepackage{xcolor}
\usepackage[urlcolor=blue]{hyperref}      % links to citations
\hypersetup{
    colorlinks = true,                    % text and not border
    citecolor = {blue},
    linkcolor = {purple},
           }
\begin{document}	
	
\title{Efficient Simulation of Quantum Many-body Thermodynamics by Tailoring Zero-temperature Tensor Network}

\author{Ding-Zu Wang}
\affiliation{School of Physics, Beihang University,100191,Beijing, China}

\author{Guo-Feng Zhang}
\email[Corresponding author: ]{gf1978zhang@buaa.edu.cn}
\affiliation{School of Physics, Beihang University,100191,Beijing, China}

\author{Maciej Lewenstein}
\email[Corresponding author: ]{maciej.lewenstein@icfo.eu}
\affiliation{ICFO - Institut de Ciencies Fotoniques, The Barcelona Institute of Science and Technology, Av. Carl Friedrich Gauss 3, 08860 Castelldefels (Barcelona), Spain}
\affiliation{ICREA, Pg. Llu\'is Companys 23, 08010 Barcelona, Spain}

\author{Shi-Ju Ran}
\email[Corresponding author: ]{sjran@cnu.edu.cn}
\affiliation{Department of Physics, Capital Normal University, Beijing 100048, China}
\date{\today}

\begin{abstract}
	Numerical annealing and renormalization group have conceived various successful approaches to study the thermodynamics of strongly-correlated systems where perturbation or expansion theories fail to work. As the process of lowering the temperatures is usually involved in different manners, these approaches in general become much less efficient or accurate at the low temperatures. In this work, we propose to access the finite-temperature properties from the tensor network (TN) representing the zero-temperature partition function. We propose to ``scissor'' a finite part from such an infinite-size TN, and ``stitch'' it to possess the periodic boundary condition along the imaginary-time direction. We dub this approach as TN tailoring. Exceptional accuracy is achieved with a fine-tune process, surpassing the previous methods including the linearized tensor renormalization group [\href{https://link.aps.org/doi/10.1103/PhysRevLett.106.127202}{Phys. Rev. Lett. 106, 127202 (2011)}], continuous matrix product operator [\href{https://link.aps.org/doi/10.1103/PhysRevLett.125.170604}{Phys. Rev. Lett. 125, 170604 (2020)}], and etc. High efficiency is demonstrated, where the time cost is nearly independent of the target temperature including the extremely-low temperatures. The proposed idea can be extended to higher-dimensional systems of bosons and fermions. 
\end{abstract}

\maketitle

{\it Introduction.---}
In recent decades, the tensor network (TN) approaches have achieved great success in the investigations of strongly-correlated quantum many-body systems. Each leap improvement of precision is generally accompanied with fundamental progresses in the mathematical or physical level. For instance, the density-matrix renormalization group (DMRG)~\cite{PhysRevLett.69.2863, PhysRevB.48.10345} better considers boundary effects than Wilson's numerical renormalization group approach (RG)~\cite{PhysRevB.21.1003}, and achieves extraordinary accuracy on simulating the ground states of one-dimensional strongly-correlated systems. It was realized afterwards that the success of DMRG relies on the matrix product state (MPS) representation as an efficient state ansatz~\cite{PhysRevLett.75.3537, RevModPhys.77.259, RevModPhys.93.045003}. The time-evolving block decimation~\cite{PhysRevLett.91.147902, PhysRevLett.93.040502} achieves the ground states by incorporating with the idea of numerical annealing, and have be successfully generalized to higher-dimensional quantum lattice models particularly those of infinite size and with translational invariance~\cite{PhysRevLett.101.250602, 2017NatCo...8.1291K, PhysRevX.8.031031}.

Simulating the finite-temperature properties is an equally important but much more challenging issue, where significant progresses have been achieved with TN~\cite{Bursill_1996, PhysRevB.56.5061, *PhysRevB.58.9142, 1997JPSJ...66.2221S, PhysRevLett.93.207204, PhysRevLett.93.207205, PhysRevB.72.220401, PhysRevLett.102.190601, 2010NJPh...12e5026S, PhysRevLett.106.127202, PhysRevB.86.045139, PhysRevB.86.245101, *PhysRevB.92.035152, *PhysRevB.99.245107, PhysRevX.8.031031, PhysRevB.95.161104, *PhysRevX.8.031082, PhysRevLett.122.070502, PhysRevB.99.205132, 2019arXiv191003329C, PhysRevB.101.195119, PhysRevB.101.220409, PhysRevLett.125.170604}. However, most methods follow the idea of annealing or RG, such as the transfer-matrix renormalization group~\cite{PhysRevB.56.5061}, linearized~\cite{PhysRevLett.106.127202} and differentiable~\cite{PhysRevB.101.220409} tensor RG methods. The process of lowering the temperatures from the infinitely-high temperature is involved in different forms. Consequently, the simulations at the extremely low temperatures, even for the simplest models such as quantum Ising and Heisenberg chains, inevitably become much less efficient or accurate. Therefore, how to efficiently calculate many-body thermodynamics, especially at the extremely low temperatures, is still an important and unsettled problem.

\begin{figure*}[tbp]
	\centering
	\includegraphics[angle=0,width=1\linewidth]{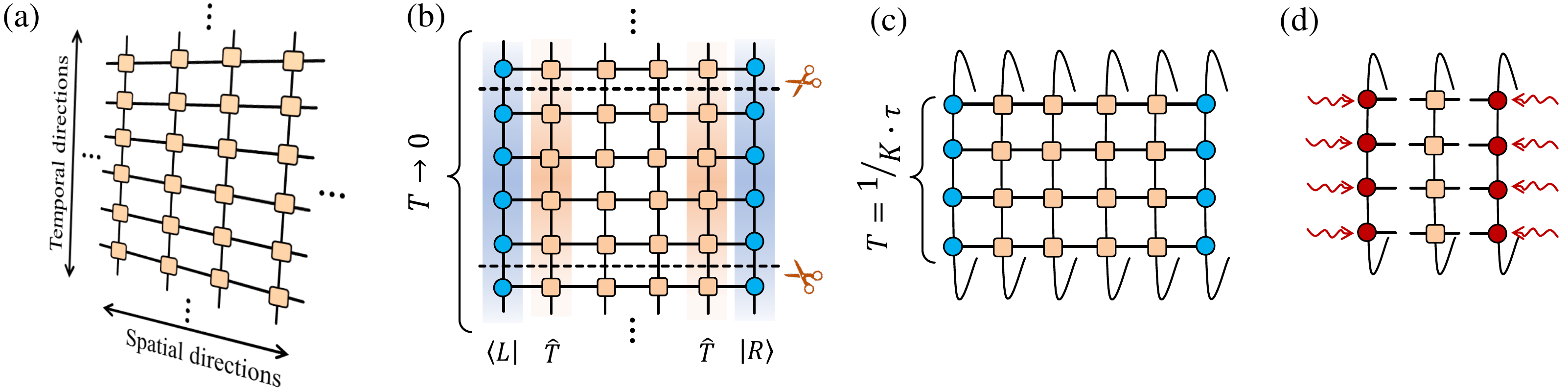}
	\caption{(Color online) In (a) we show the infinite square tensor network representing the density matrix $e^{-\hat{H}/T}$ of the quantum system at $T \to 0$ in the thermodynamic limit. In (b) we show the boundary MPS's ($\langle L|$ and $|R\rangle$) of the infinite TN in (a) along temporal (vertical) direction. Then scissoring $K$ layers from the TN along the imaginary-time direction. (c) A finite part with $K$ layers is scissored from the infinite TN in (b), and stitched to possess the periodic boundary condition alone the temporal direction. It approximately represents the finite-temperature reduced density operator with $T=1 / K\cdot\tau$. In (d) we illustrate $\langle L|\hat{T}|R \rangle$ [Eq. (\ref{app-eq-f})] for fine-tuning.}
	\label{Fig1}
\end{figure*}

To this aim, we propose to access the finite-temperature properties by ``tailoring'' the zero-temperature TN, instead of lowering the temperature from infinite. By ``zero-temperature TN'' we mean the infinite-size TN whose contraction gives the zero-temperature density matrix. In specific, we scissor a finite part of the TN with the boundary MPS~\cite{PhysRevLett.75.3537, 1992CMaPh.144..443F} along the temperature (or imaginary time), and stitch it to possess the periodic boundary condition for evaluating observables [Fig. \ref{Fig1}]. The height of the scissored TN determines the temperature we are aiming at. The boundary MPS's of the infinite TN can be efficiently obtained by the ground-state algorithms~\cite{PhysRevLett.69.2863, PhysRevB.48.10345, PhysRevLett.91.147902, PhysRevLett.93.040502, PhysRevB.80.094403, PhysRevE.93.053310, PhysRevB.96.155120}. Additionally with a fine-tuning process to optimize the boundary MPS in the case of the finite height, exceptional accuracy is achieved, significantly surpassing the state-of-the-art methods including linearized~\cite{PhysRevLett.106.127202} and differentiable tensor renormalization group~\cite{PhysRevB.101.220409} (LTRG and $\partial$TRG in short), and continuous matrix product operator (cMPO)~\cite{PhysRevLett.125.170604}. TN tailoring could access any temperature with superior efficiency. Impressively, our results show that the time cost is nearly independent of the target temperature including the extremely-low temperatures. Our work can be generalized to simulate interacting bosons and fermions in one and higher dimensions, and quantum fields in the continuous space.

{\it Tensor network tailoring.---}
Let us start with the Hamiltonian of one-dimensional (1D) quantum many-body system given by 
\begin{eqnarray}\label{app-eq-H}
	\hat{H} = \sum_{n=1}^{N}\hat{h}_{n,n+1},
\end{eqnarray}
where the thermodynamic limit $N \to \infty$ is taken. The density operator at the temperature $T$ can be presented as
\begin{eqnarray}\label{app-eq-Z}
	&&\hat{\rho}(T) = e^{-\beta \hat{H}} = (e^{-\tau \hat{H}})^{K},
\end{eqnarray} 
with the Trotter slice $\tau$ a small positive constant and the inverse temperature (or imaginary time) $\beta = 1 / T = \tau K$. The Plank and Boltzmann constants are set as one. The second-order Trotter-Suzuki decomposition is applied as
\begin{equation}\label{app-eq-TS}
	e^{-\tau \hat{H}} \simeq e^{- \frac{\tau}{2}\sum_{\text{odd }n} \hat{h}_{n,n+1}} e^{-\tau \sum_{\text{even }n} \hat{h}_{n,n+1}} e^{- \frac{\tau}{2}\sum_{\text{odd }n}\hat{h}_{n,n+1}}.
\end{equation} 
Then $\hat{\rho}(T)$ can be represented by an TN that is formed by the fourth-order tensors
\begin{eqnarray}\label{app-eq-local}
	U_{s_{1}s_{2},s'_{1}s'_{2}}^{[\tau]} = \langle s_{1}s_{2}| e^{-\tau \hat{h}_{n,n+1}} |s'_{1}s'_{2}\rangle, \\
	U_{s_{1}s_{2},s'_{1}s'_{2}}^{[\frac{\tau}{2}]} = \langle s_{1}s_{2}| e^{-\frac{\tau}{2} \hat{h}_{n,n+1}} |s'_{1}s'_{2}\rangle,
\end{eqnarray} 
where $\{|s_{i}\rangle\}$ and $\{|s'_{i}\rangle\}$ denote the orthonormal basis of spins. To proceed, we decompose $U^{[\tau]}$ by singular-value decomposition as
\begin{eqnarray}\label{app-SVD}
	U_{s_{1}s_{2},s'_{1}s'_{2}}^{[\tau]}
	&& = \sum_{\alpha} V^{L}_{s_{1}s'_{1},\alpha} S_{\alpha} V^{R}_{s_{2}s'_{2},\alpha}
\end{eqnarray} 
with $S$ the singular-value spectra. For convenience, we then transfer TN so that it contains the infinite copies of one tensor. We have a square TN formed by one inequivalent tensor 
\begin{eqnarray}
	&& \Theta_{\alpha,s_{3}'s_{3},s_{4}s'_{4},\beta} \nonumber \\
	&& = \sum_{s_{1},s'_{1},s_2,s'_{2}} U^{[\frac{\tau}{2}]}_{s_3,s'_{3},s_2,s_1} \tilde{V}^{L}_{s_{1}, s'_{1}, \beta} \tilde{V}^{R}_{s_{2}, s'_{2}, \alpha} U^{[\frac{\tau}{2}]}_{s'_{2},s'_{1},s_4,s'_{4}}.
\end{eqnarray}  
with $\tilde{V}^{L}_{s_{1}, s'_{1}, \alpha} = \sum_{\alpha} V^{L}_{s_{1}, s'_{1}, \alpha} \sqrt{S}_{\alpha}$ and $\tilde{V}^{R}_{s_{2}, s'_{2}, \alpha}=\sum_{\alpha}\sqrt{S}_{\alpha} V^{R}_{s_{2}, s'_{2}, \alpha}$.

Taking $T \to 0$ (or $K \to \infty$), $\hat{\rho}(T)$ becomes the projector that projects the initial state to the ground state. The corresponding TN is infinite along both the spatial and imaginary-time directions [Fig. \ref{Fig1}(a)]. The ground-state properties can be efficiently and accurately obtained by calculating the boundary MPS's $|L\rangle$ and $|R\rangle$ using, e.g., infinite time-evolving block decimation~\cite{PhysRevLett.98.070201} and TN encoding~\cite{PhysRevE.93.053310}. $|L\rangle$ and $|R\rangle$ are the left and right dominant eigenvectors of the transfer matrix product operator $\hat{T}$, satisfying $\lambda^{\ast} |L\rangle = \hat{T}^{\dagger} |L\rangle$ and $\lambda  |R\rangle = \hat{T} |R\rangle$ (suppose there exists the unique dominant eigenvalue $\lambda$) [Fig. \ref{Fig1}(b)].  Without losing generality, we assume $|L\rangle$ and $|R\rangle$ to be uniform MPS's~\cite{1992CMaPh.144..443F}. 

Our key idea is scissoring $K$ layers from the TN along the imaginary-time direction [Fig. \ref{Fig1}(b)]. The scissored TN approximately represents the finite-temperature density operator at $T=1/(K\tau)$ [Eq. \ref{app-eq-Z}]. The thermodynamic quantities, such as partition function $Z = \text{Tr}(\hat{\rho})$ [Fig. \ref{Fig1}(c)] and energy $E = \text{Tr}(\hat{\rho} \hat{H}) / Z$, can be reached by stitching the TN to the periodic boundary condition along the imaginary-time direction. From the perspective of quantum simulation, effective Hamiltonians can be defined from the scissored TN and boundary MPS's, which reduces the finite-size effects~\cite{PhysRevB.99.205132}

Our approach is essentially different from the existing algorithms. Taking the transfer-matrix renormalization group as an example, it utilizes density matrix renormalization group to calculate the boundary MPS that represents the RG along the imaginary time. The number of the RG steps increases linearly with $K$. The LTRG achieves higher accuracy and stability than the transfer-matrix renormalization group by employing the idea of numerical annealing and the matrix product operator (MPO) representation. The number of the annealing steps also increases linearly with $K$. In addition, the error will in general accumulate as $K$ increases.

Another key step of our approach is fine-tuning, which further increases the accuracy of our method. Starting from scissored TN and the boundary MPS's, we optimize the tensors in the boundary MPS's by maximizing the free energy per site 
\begin{equation}\label{app-eq-f}
	f  = \max_{A, B} \left[ -\frac{1}{\beta} \Big (\ln \langle L|\hat{T}|R\rangle - \ln \langle L|R\rangle\Big) \right].
\end{equation} 
with $A$ and $B$ the inequivalent tensors of $|L\rangle$ and $|R\rangle$, respectively [Fig. \ref{Fig1} (d)]. $A$ and $B$ are updated using gradient descent
\begin{eqnarray}\label{app-eq-grad}
	A \leftarrow A + \eta \frac{\partial f}{\partial A}, \\
	B \leftarrow B + \eta \frac{\partial f}{\partial B},
\end{eqnarray}
with $\eta$ the gradient step or learning rate. The gradients are obtained using the automatic differentiation technique of TN~\cite{PhysRevX.9.031041,pytorch}. Supposing $f$ reaches the maximum, one would have the eigenvalue
\begin{eqnarray}\label{app-Z}
	\lambda = \frac{\langle L| \hat{T} |R\rangle}{\langle L | R \rangle},
\end{eqnarray} 
and the partition function $Z = \text{Tr}(e^{-\beta \hat{H}}) = \lambda^N$ with the number of spins $N \to \infty$. The fine-tuning process optimizes the boundary MPS's to the global optimum for the target temperature, while the translational invariance and periodic boundary condition of the boundary MPS's are respected.

\begin{figure}[tbp]
	\centering
	\includegraphics[angle=0,width=0.9\linewidth]{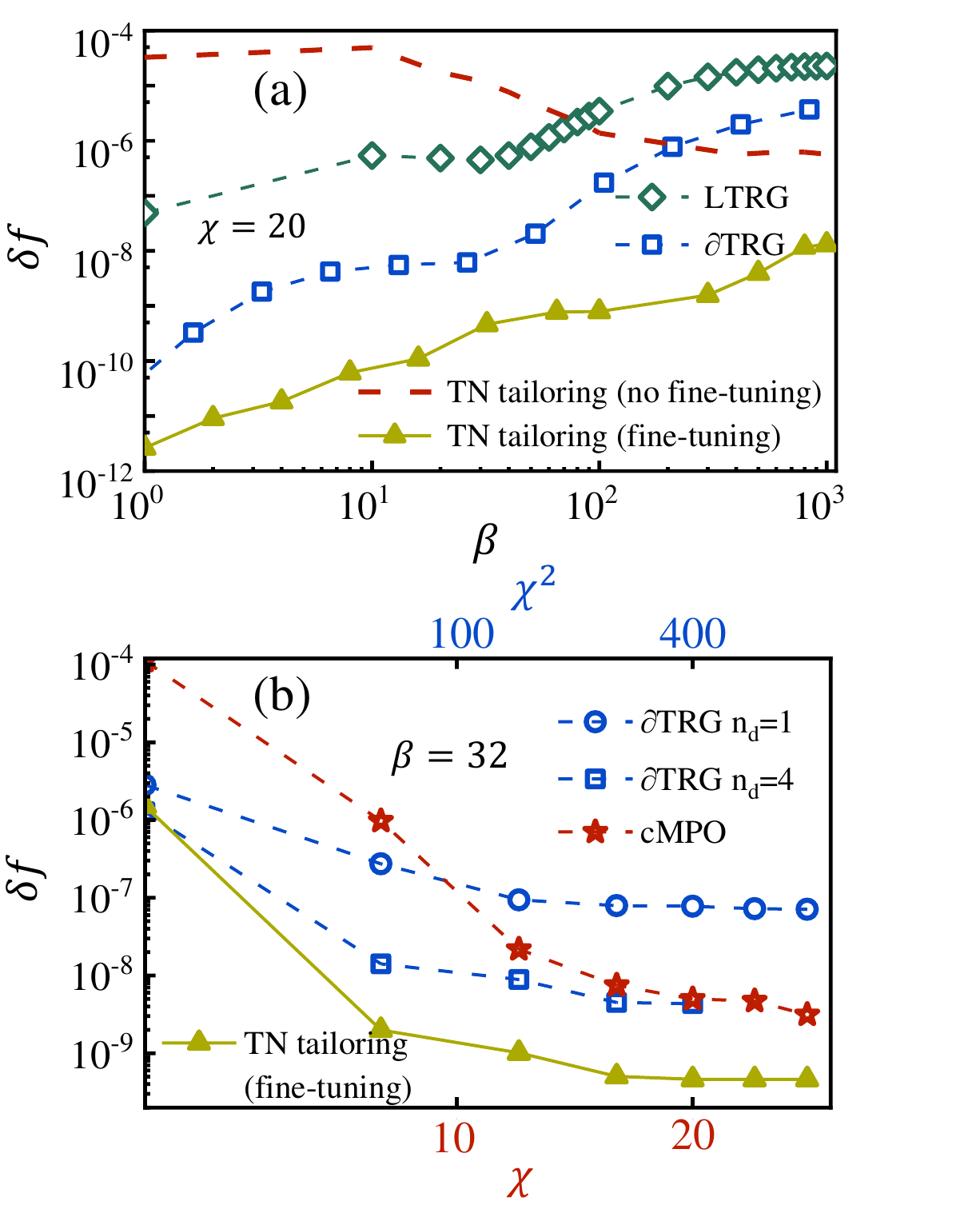}
	\caption{(Color online) The relative errors of the free energy $\delta f$ [Eq. (\ref{eq-df})] obtained by LRTG, $\partial$TRG, cMPO, and TN tailoring with or without fine-tuning on the critical quantum Ising chain. In (a) we show $\delta f$ versus the inverse temperature $\beta$. For $\partial$TRG, we take the optimization depth $n_{d}=4$ and overall sweep iterations $n_{s}=3$. In (b) we show $\delta f$ versus the bond dimension $\chi$ (bottom, for cMPO and TN tailoring) or $\chi^{2}$ (top, for $\partial$TRG) at $\beta=32$.}
	\label{Fig2}
\end{figure}

{\it Results and comparisons.---}
The performance of TN tailoring is tested on the infinite quantum Ising chain
$\hat{H} = \sum_{n} \hat{S}_n^x \hat{S}_{n+1}^x - h \sum_{n} \hat{S}^z_n$. The relative errors of the free energy per site is given by
\begin{eqnarray} \label{eq-df}
	\delta f =\frac{|f-f_{\text{exact}}|}{|f_{\text{exact}}|} ,
\end{eqnarray}
with $f_{\text{exact}}$ the analytic solution~\cite{1985issm.book....5B}. We test the approaches at the critical field $h=0.5$. 

Fig. \ref{Fig2}(a) shows the $\delta f$ obtained by the existing algorithms, including LTRG~\cite{PhysRevLett.106.127202}, $\partial$TRG~\cite{PhysRevB.101.220409}, and cMPS~\cite{PhysRevLett.125.170604} at different inverse temperatures $\beta = 1/T$. Without fine-tuning, lower $\delta f$ than LTRG and $\partial$TRG is achieved at the low temperatures. For higher temperatures, $\delta f$ becomes larger since the TN and boundary MPS's are scissored from those at zero temperature. Nevertheless, we still have $\delta f \sim O(10^{-4})$ or lower.

With the fine-tuning process, $\delta f$ is dramatically decreased at all temperatures from $T \sim O(1)$ to $O(10^{-3})$. It is several orders of magnitude lower than that obtained by LTRG and $\partial$TRG. An important fact is that $\delta f$ does not increase with $T$ (i.e., decrease with $\beta$). It means that any temperatures can be accurately reached from zero temperature in the fine-tuning process. As $T$ lowers, $\delta f$ increases due to the growth of the correlations (or entanglement) of the system.

Fig. \ref{Fig2}(b) compares $\delta f$ with different bond dimensions $\chi$ of the MPS or MPO at a low temperature $T=1/32$. We show the results for the TN tailoring and cMPO approaches against $\chi$, and $\partial$TRG against $\chi^{2}$. This is because $\chi$ in the cMPO and TN tailoring approaches refers to the bond dimension of the boundary MPS. Consequently, the complexity scales approximately with $\chi^6$. For the $\partial$TRG, $\chi$ refers to the bond dimension of the MPO, thus the complexity scales as $\chi^4$. Our method achieves the highest accuracy among all these approaches. For instance, we have $\delta f \simeq 4.0\times10^{-10}$ and $4.9\times10^{-9}$ for $\chi=20$ using TN tailoring and cMPO, respectively, and $\delta f \simeq 4.3\times10^{-9}$ for $\chi=400$ using $\partial$TRG~\cite{dTRG}. 

Notably, the boundary MPS's in our scheme can be readily written in the continuous MPS form, in order to respect the continuous nature of the temperature as the cMPO does. However, this will not significantly affect the accuracy, since the key of our high accuracy lies in the tailoring (i.e., scissoring and stitching from the zero-temperature TN) but not the continuity. 

\begin{figure}[tbp]
	\centering
	\includegraphics[angle=0,width=0.95\linewidth]{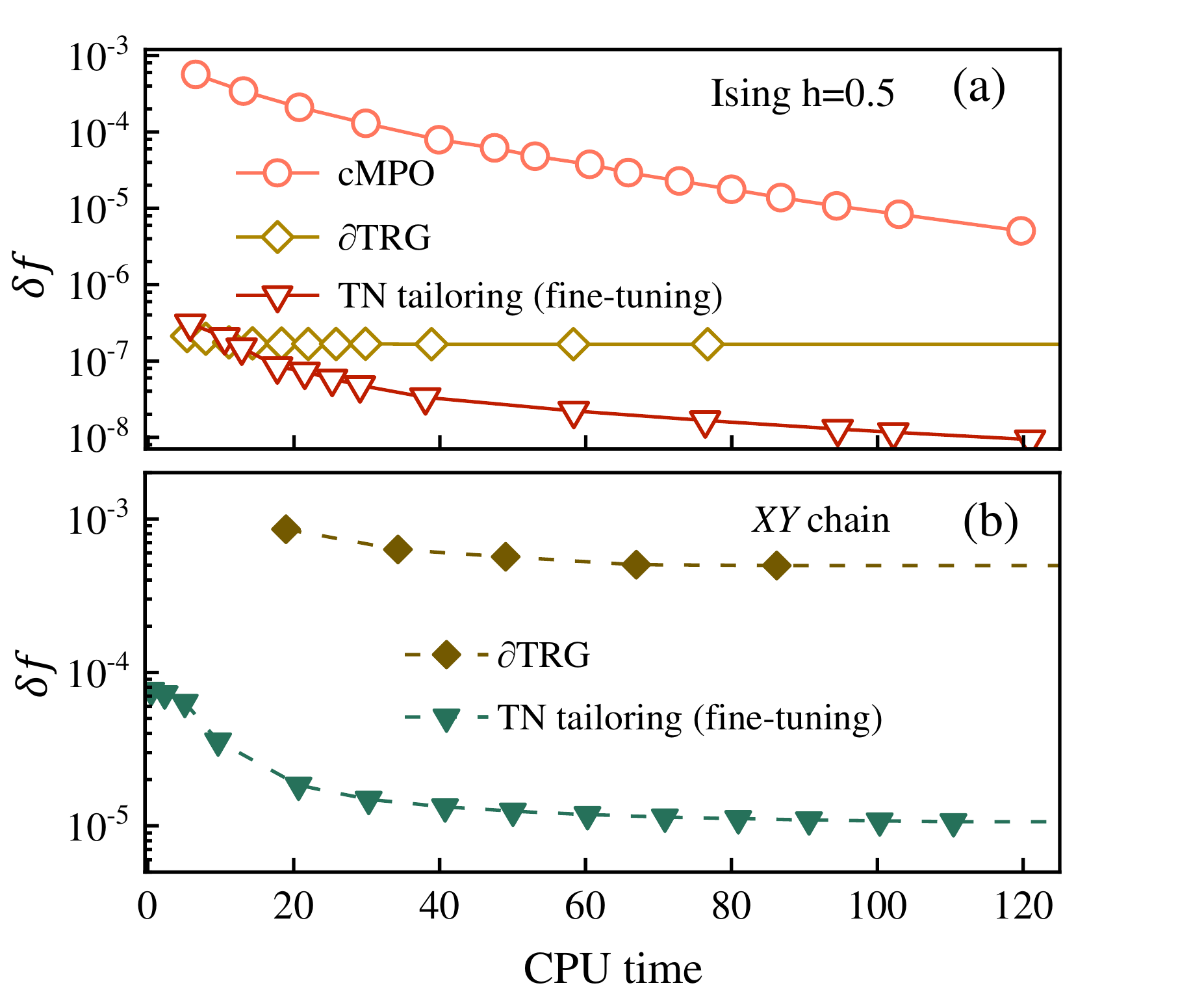}
	\caption{(Color online) In (a) we show the relative errors $\delta f$ versus CPU time by $\partial$TRG, cMPO, and TN tailoring with fine-tuning on the critical Ising chain. In (b), the results on the {\it XY} chain are demonstrated. We fixed $\beta\backsimeq100$, the bond dimension $\chi=20$, Trotter slice $\tau=10^{-4}$, and the learning rate [Eq. (\ref{app-eq-grad})] in the fine-tuning of TN tailoring $\eta=10^{-9}$. The optimization depth and overall sweep in $\partial$TRG are initially taken as $n_{d}=1$ and $n_{s}=3$, and then increase them until the accuracy converges.}
	\label{Fig3} 
\end{figure}

To fairly compare the efficiency, in Fig. \ref{Fig3} we further demonstrate $\delta f$ versus the CPU-time costs using different methods. Various time costs and $\delta f$ are achieved by varying the optimization parameters. The time cost of TN tailoring is the summation of the time for obtaining the zero-temperature TN and that for the fine-tuning process. All tasks are run on the same computer~\cite{time}. In Fig. \ref{Fig3}(a), we show the relative errors of the quantum Ising chain ($h=0.5$). The cMPO approach gives much higher $\delta f$ with a fixed time cost. For $\partial$TRG, the obtained $\delta f$ is much lower than that by cMPO. However, increasing time cost will just lead to minor improvement on the accuracy. With TN tailoring, $\delta f$ decreases with the time cost, and becomes much lower than those obtained by cMPO and $\partial$TRG. We also test on the {\it XY} chain with the Hamiltonian $\hat{H} = \sum_{n} \hat{S}_n^x \hat{S}_{n+1}^x + \hat{S}_n^y \hat{S}_{n+1}^y$, shown in Fig. \ref{Fig3}(b). TN tailoring achieves much lower $\delta f$ in the whole range of the time cost.

\begin{figure}[tbp]
	\centering
	\includegraphics[angle=0,width=1\linewidth]{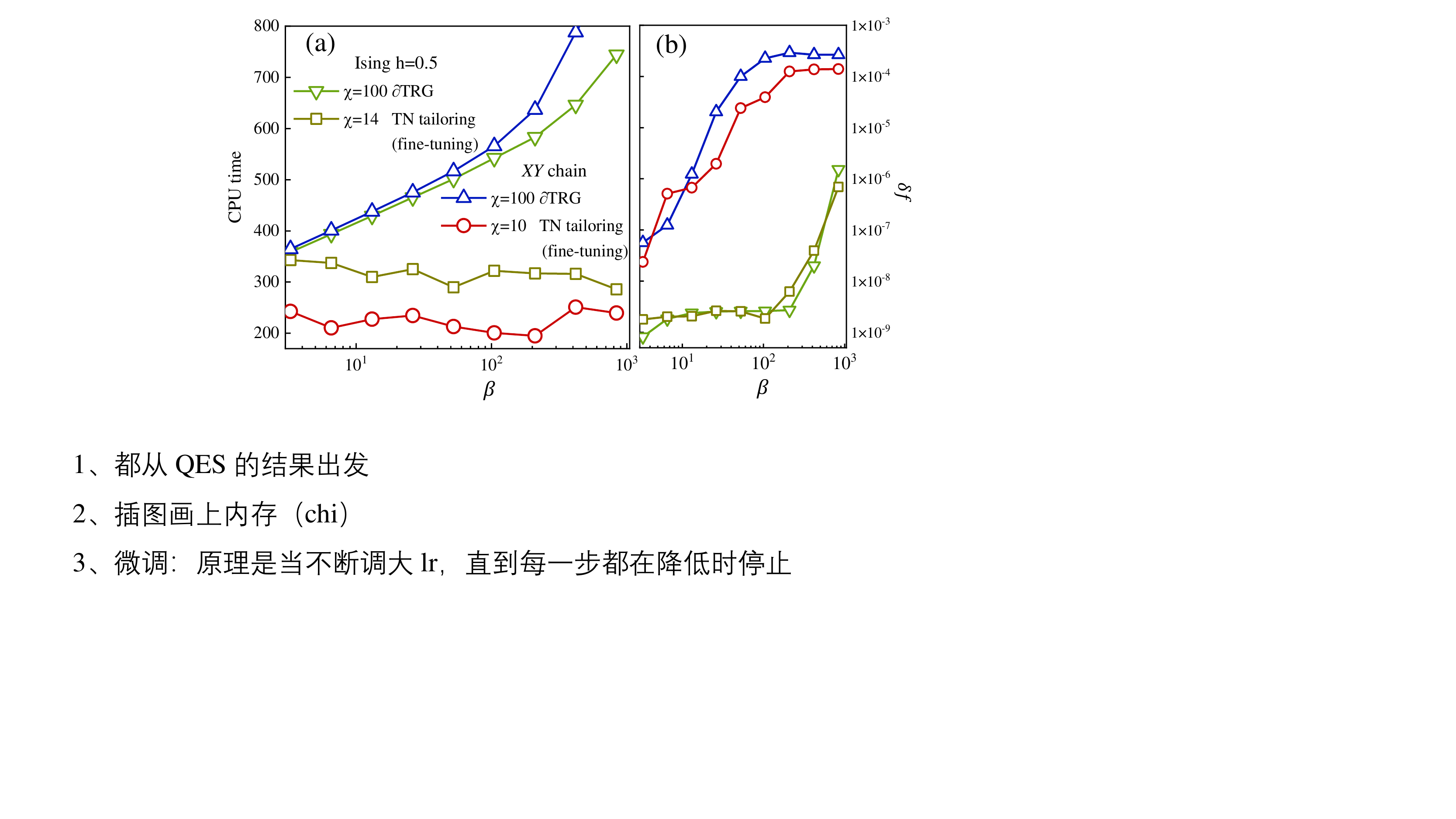}
	\caption{(Color online) (a) The cost of CPU time versus $\beta$ on the critical Ising and {\it XY} chains. In (b) we show the relative errors $\delta f$ from the data in (a). The bond dimensions are taken in such a way that the two approaches give similar accuracies. We take the leaning rate as $\eta\backsimeq2.0\times10^{-9}$, and the optimization depth and overall sweep in $\partial$TRG as $n_{d}=4, n_s=3$, respectively.}
	\label{Fig4}
\end{figure}

A unique advantage of TN tailoring is that its time cost is almost independent of the target temperature, including the extremely-low temperatures. To provide a quantitative demonstration, we show the time costs of TN tailoring and $\partial$TRG at different temperatures for the quantum Ising and XY chains [Fig. \ref{Fig4}(a)], keeping the errors $\delta f$ of these two approaches close to each other to make the comparison fair [Fig. \ref{Fig4}(b)]. The time cost of $\partial$TRG apparently increases with the inverse temperature $\beta$. In contrast, for TN tailoring, the time cost keeps nearly constant in the whole temperature range.

{\it Summary and Discussion.---}
Different from the frequently-used annealing or renormalization-group treatments, we here propose the TN tailoring approach that accesses the finite-temperature properties by ``tailoring'' an infinite-size TN that represents the zero-temperature density operator. With a fine-tuning process, superior accuracy is achieved, surpassing the existing powerful methods including the linearized tensor renormalization group, continuous matrix product operator, and etc. The efficiency of TN tailoring is also demonstrated. Its time cost is almost independent  the target temperatures including the extremely-low temperatures. TN tailoring provides a novel approach to simulate interacting bosons and fermions in higher dimensional, as well as the quantum fields in the continuous space.

\AtEndEnvironment{thebibliography}{
	\bibitem{pytorch} The official website of PyTorch is at \url{https://pytorch.org/}
	
	\bibitem{dTRG} The $\partial$TRG shows different performances depending on the depth $n_{d}$. Its performance improves as $n_{d}$ increases. Here we follow the papers about $\partial$TRG and the supplemental material of cMPO with depth $n_{d}=4$. The LTRG is already surpassed by the cheapest choice $n_{d}=1$ in $\partial$TRG. For $n_{d}=4$ the precision is converged up to the bond dimension 400. With a second-order Trotter decomposition, $\partial$TRG it has a Trotter error of the order $10^{-10}$, which is negligible.
	
	\bibitem{time} The calculations are carried out on the same desktop with Intel Core i9-10900K CPU.}
	
%For TN tailoring, after determining the optimizing parameters $\eta$, the time cost is only related to the number of optimized steps. In the fine-tune process, the more steps the higher the accuracy, and the cost time is increasing correspondingly. The cMPO algorithm is similar. For $\partial$TRG, the different parameters $n_{d}, n_{s}$ directly determine the cost time and accuracy, here our strategy is to start with the optimizing parameter $n_{d}=1, n_{s}=3$, first increase $n_{d}$ to 8, and then increase $n_{s}$ to 8.

\begin{acknowledgments}
	The authors are grateful to Gang Su, Wei Li, Han Li, Kai Xu, Han-Jie Zhu, Bin-Bin Chen, Hao Zhu and Leticia Tarruell for stimulating discussions. This work is supported by NSFC (No. 12004266, No. 11834014 and Grant No. 12074027), Beijing Natural Science Foundation (Grant No. 1192005 and No. Z180013), Foundation of Beijing Education Committees (No. KM202010028013), and the key research project of Academy for Multidisciplinary Studies, Capital Normal University. M.L. acknowledges support from ERC AdG NOQIA, Agencia Estatal de Investigación (“Severo Ochoa” Center of Excellence CEX2019-000910-S, Plan National FIDEUA PID2019-106901GB-I00/10.13039/501100011033, FPI), Fundació Privada Cellex, Fundació Mir-Puig, and from Generalitat de Catalunya (AGAUR Grant No. 2017 SGR 1341, CERCA program, QuantumCAT U16-011424 , co-funded by ERDF Operational Program of Catalonia 2014-2020), MINECO-EU QUANTERA MAQS (funded by State Research Agency (AEI) PCI2019-111828-2/10.13039/501100011033), EU Horizon 2020 FET-OPEN OPTOLogic (Grant No 899794), and the National Science Centre, Poland-Symfonia Grant No. 2016/20/W/ST4/00314, Marie Sklodowska-Curie Grant STRETCH No. 101029393.
\end{acknowledgments}

\bibliography{Primary_manuscript}

%\clearpage

%\section*{SUPPLEMENTAL MATERIAL}

%\appendix

%\section{aa}
% some text...
%\section{Some Examples 2}
% some text...

\end{document}